\documentclass[11pt,twoside]{article}
\usepackage{asp2010}

\resetcounters

\begin{document}

\title{The Evolution of the Large-scale ISM: Bubbles, Superbubbles and Non-Equilibrium Ionization}
\author{Miguel A. de Avillez$^1$, Dieter Breitschwerdt$^2$
\affil{$^1$Department of Mathematics, University of \'Evora, R. Rom\~ao Ramalho 59, 7000 \'Evora, Portugal}
\affil{$^2$Zentrum f\"ur Astronomie und Astrophysik, Technische Universit\"at Berlin, Hardenbergstr.~36, D-10623 Berlin,
Germany}
}

\begin{abstract}
The ISM, powered by SNe, is turbulent and permeated by a magnetic field (with a mean and a turbulent component). It
constitutes a frothy medium that is mostly out of equilibrium and is ram pressure dominated on most of the temperature
ranges, except for T$< 200$ K and $T> 10^6$K, where magnetic and thermal pressures dominate, respectively. Such lack of
equilibrium is also imposed by the feedback of the radiative processes into the ISM flow. Many models of the ISM or
isolated phenomena, such as bubbles, superbubbles, clouds evolution, etc., take for granted that the flow is in the
so-called collisional ionization equilibrium (CIE). However, recombination time scales of most of the ions below
$10^{6}$ K are longer than the cooling time scale. This implies that the recombination lags behind and the plasma is
overionized while it cools. As a consequence cooling deviates from CIE. This has severe implications on the
evolution of the ISM flow and its ionization structure. Here, besides reviewing several models of the ISM, including
bubbles and superbubbles, the validity of the CIE approximation is discussed, and a presentation of recent developments
in modeling the ISM by taking into account the time-dependent ionization structure of the flow in a full-blown numerical
3D high resolution simulation is presented.
\end{abstract}

\section{Introduction}

In the last decade due to the substantial increase of computing power, sophisticated models of the interstellar medium
(ISM), including the disk-halo interaction, bubble and superbubble evolution, molecular clouds evolution and
fragmentation, formation of shock compressed layers, etc., in a turbulent supernova-driven medium, have been developed.
In general these models included cooling and heating, which determine the physical state of the gas. Interstellar
cooling can be the result of line and continuum plasma emission processes, as well as of adiabatic expansion of
over-pressured gas. The importance of the former depends on the amount of atoms and ions present in the flow, whereas
the latter is related to thermodynamical processes. For optically thin interstellar plasmas, frequently the assumption
of collisional ionization equilibrium (CIE) is used, according to which the collisional excitation of the gas is
followed by photon emission, with the number of ionizations being equal to the number of (dielectronic or radiative)
recombinations. However, CIE violates \emph{detailed balancing}, since collisional excitation involves three particles,
an atom or ion, an electron to collide with, and a second electron which is ejected, whereas in radiative recombination,
the third particle is a photon, leaving the system. Thus, strictly speaking, CIE can never be maintained in an evolving
plasma, although it might be a fair approximation, especially in hot (T$>10^{6}$ K) environments. For lower temperatures
recombinations are not synchronized with the cooling and therefore, deviations from CIE inevitably occur
\citep[see e.g.,][]{kafatos1973, shapiro1976, st1993}.

The CIE assumption in all ranges of temperatures, or at least in part of it, in the models discussed below implies the
use of a specific cooling function, for given abundances, in the computational domain and at each time step. Obviously,
the history of the ionization structure, which depends on the thermodynamic path of the plasma, will be lost, in
contrast to a full non-equilibrium ionization (NEI) structure giving rise to a \emph{time-dependent} cooling function,
which is quite different than that obtained under CIE conditions. 

The structure of this paper is the following: In Section 2 a review on ISM modelling including superbubbles and
disk-halo interaction, is presented. Sections 3 and 4 deal with the collisional ionization conditions adopted in the
models, and compare CIE with NEI results, respectively. As cooling is a process that depends on the history of the
plasma, we discuss in Section 5 the cooling functions and emission spectra of gas in the latest simulations of the ISM.
The paper is closed with a final remarks and conclusions section.

\section{From the 3-Phase Model to Present day ISM and Superbubbles Models}

Theoretical studies during the last three decades have culminated in the supernova regulated Interstellar medium models
of \citet{cs74} and in the widely accepted ``standard-model'' \citep{mo1977}, in which the gas is distributed into three
phases in global pressure equilibrium, a cold and warm neutral phase (CNM and WNM, respectively), a warm ionized (WIM)
and a hot intercloud (HIM) medium. There is global mass balance by evaporation, ionization and condensation, and energy
balance between supernova (SN) energy injection and radiative cooling. Most of the Galactic volume (up to 70-80\%) is
filled with the hot ($>10^{5}$ K) low-density ($\sim 10^{-3}$ cm$^{-3}$) ionized gas, interspersed by cold neutral and
relatively dense clouds. Observationally it was difficult to determine reliable volume filling factors of the hot phase
in our Galaxy due to the observational vantage point. However, it became evident that the small surface coverage of
H{\sc i} holes in external galaxies \citep[e.g.,][]{brinks1986} argues for a much lower volume filling factor of the hot
phase there.

Furthermore, Population I stars typically born in OB associations \citep[see, e.g.,][]{ms1979,mk1987}. Since more
massive stars evolve more rapidly than low mass stars, the associations still exist when the first SNe occur.
Hence, SNe are not randomly distributed over the entire galaxy. As a result of their motions, up to 40\% of the stars
explode in the field \citep[see review by][]{ferriere2001}. Therefore, the amount of hot gas will mostly be
concentrated in isolated pools (forming superbubbles), which may burst through the thick disk (composed of H{\sc i}
\citep{lockman1984} and H{\sc ii} \citep{reynolds1985} layers) into the halo like \emph{chimney funnels} \citep{ni1989}.
The superbubble evolution in a static density stratified environment was modelled hydrodynamically in two dimensions by,
e.g., \citet{tomisaka1986}, \citet{mlmn1989}, and \citet{tt1990}. \citet{tomisaka1998} and \citet{stil2009} developed
three-dimensional simulations of a superbubble evolving in a static magnetized medium having a stratified structure
along the direction perpendicular to the Galactic midplane.

With the presence of such an extended thick disk, arguments were put forward that the break-out of bubbles
and superbubbles could be inhibited (unless they occurred at a significant height above the disk) and only the most
energetic superbubbles (SBs) would achieve blow-out of the disk \citep[see][]{km1992}. Things might even become worse
as superbubble (SB) break-out may be inhibited by a large-scale disk parallel magnetic field
\citep[e.g.][]{mineshige1993, tomisaka1998}. 
On the other hand, it seems paradoxical that terrestrial plasmas are so hard to confine, whereas ISM plasmas should be
magnetically tied down to the disk. In fact in both cases small and large scale instabilities destroy the highly
symmetric configuration, like e.g. the Parker (1966) instability in the astrophysical context. Observationally, owing to
the high sensitivity and large throughput of the ROSAT XRT \citep{trumper1983}) and its PSPC instrument,
a number of normal spiral galaxies with soft X-ray halos were detected, e.g. NGC 891 \citep{brp1994}
and NGC 4631 \citep[see][]{vogler1996}. In some cases, even local correlations between H$\alpha$, radio continuum and
soft X-rays were found \citep{dettmar1992}, arguing for local outflows as it had been suggested in the Galactic fountain
\citep{bregman1980,kahn1981}, chimney \citep{ni1989} and the Galactic wind model \citep{breimv1991, bs1994}.

These models capture some of the structure but not all of the essential physics. Taking into account that the ISM is a
turbulent and compressible system (von Weizs\"acker 1951), in which cooling and heating determine the physical state of
the gas, more complex and sophisticated galactic disk \citep[two-dimensional:][]{cp1985,cb1988,rbn1993} and
disk-halo interaction two- \citep{rb1995} and three-dimensional
\citep{korpi1999a, av2000,ab2004,ab2005,joung2006,melioli2009} models were developed. In turn, some of these models have
been used to follow the evolution of superbubbles within a realistic supernova-driven turbulent magnetized medium (in
contrast to previous simulations that used a static medium) by \citet{korpi1999b} and \citet{ab2005} (a field composed
of a random and mean components with a total strength of 4.5 $\mu$G was used) finding that blow-out is much more likely
than previously thought, mainly because bubbles evolve in an inhomogeneous background medium. However, Korpi et al.'s
simulations are limited by the usage of a small grid extending only up to 1 kpc in the direction perpendicular to the
disk in both sides of the midplane. Hence, the disk-halo-disk cycle could not be established nor followed. The
simulations of \citet{melioli2009} calculate the time evolution of H{\textsc{i}}, H{\textsc{ii}},
C{\textsc{ii}}-C{\textsc{iv}}, and O{\textsc{i}}-O{\textsc{iii}} ions at T$\leq 10^{6}$ K and determine their
contributions to the cooling function. For T$> 10^{6}$ K they use a CIE cooling curve. Such setup has severe
implications in the history and cooling of the plasma as will be discussed below.

The evolution of superbubbles can be much improved by, instead of injecting energy in a continuous way into a single
point, identifying the missing stars in an association and follow in space and time all the stars during their main
sequence life time until they explode. This methodology has been used in the disk-halo simulations, including those
related to the evolution of the Local and Loop I superbubbles in a turbulent medium \citep{ba2006, ab2009}. In this way
one can perform a detailed simulation of the properties of the bubbles and their spectroscopic observables, allowing a
direct comparison with observations. These simulations have been successful in reproducing not only the spatial
structure of the bubbles, but also the observed column densities of Li-like ions C{\textsc{iv}}, N{\textsc{v}} and
O{\textsc{vi}} and their ratios.

\section{Plasma Emission Modelling}

A further improvement in modelling the ISM comprises the full-blown non-equilibrium ionization structure (resulting
from the ten most abundant elements), where the ionization, thermal and dynamical history of the plasma are fully
nonlinearly coupled and tracked simultaneously both in space and time at high resolution. In order to achieve
this, we developed a plasma emission code (hereafter EPEC - Eborae Plasma Emission Code) which can be coupled to
any MHD software through the proper interfacing calls. EPEC is written in Fortran 2003 in an object-oriented way making
large use of procedure pointers. It is prepared to run both on multi-core CPUs (using OpenMP) as well as on NVIDIA GPUs
(graphics cards processor units) by means of CUDA Fortran.

\subsection{Abundances}

EPEC includes the ten most abundant elements in nature (H, He, C, N, O, Ne, Mg, Si, S and Fe) and the default solar
abundances are those recommended by \citet[AGSS2009]{agss2009}: $\log A(X/H)_{\odot}$= -1.07 (He), -3.57 (C),
-4.17 (N), -3.31 (O), -4.07 (Ne), -4.40 (Mg), -4.49 (Si), -4.88 (S), and -4.50 (Fe). Other abundances (for
comparison studies with previously  published results) are also available, e.g., \citet{allen1973},
\citet[AG1989]{ag1989} and \citet[GAS2007]{gas2007}. The latter are used by \citet{gs2007}, but with the Ne
overabundance ($\log \mbox{A(Ne/H)}_{\odot}=-3.71$) of \citet{dt2005}. GAS2007 and AGSS2009 propose C, N, O and Ne
abundances smaller than those recommended by AG1989, but AGSS2009 increases slightly those values from GAS2007. This
variation in the abundances results from the improvement on three-dimensional hydrodynamical solar model atmospheres,
that include a relaxation assumption on the local thermodynamic equilibrium and improvements in the atomic and molecular
data (see discussion in, e.g., AGSS2009).

\subsection{Atomic and Cooling Processes}

The adopted physical processes in this work are the electron impact ionization, excitation-autoionization,
radiative and dielectronic recombination, charge-exchange recombination and ionization reactions, continuum and line
emissions. Electron impact ionization rates fits are taken from \citet[MAT2007]{mat2007} for all H, He, C, N, O, Ne
ions, Mg{\textsc{i}}-Mg{\textsc{iii}}, Si{\textsc{i}}-Si{\textsc{viii}}, S{\textsc{i}}-S{\textsc{v}},
and Fe{\textsc{i}}-Fe{\textsc{xi}}; The experimental data of \citet{fogle2008} is used to fit the rates for
C{\textsc{iii}}, N{\textsc{iv}}, and O{\textsc{v}}. Data for the remaining ions is taken from \citet[MAZ1998]{maz1998}.
Excitation-autoionization rates are taken from MAT2007 for C{\textsc{iv}}, N{\textsc{v}}, O{\textsc{vi}},
Ne{\textsc{viii}}, Si{\textsc{iii}}-Si{\textsc{iv}}, S{\textsc{iii}}, S{\textsc{v}} and Fe{\textsc{xi}}. For the
remaining ions where EA is important we follow \citet[AR85]{ar1985}. 

Radiative recombination rates are fitted following \citet{vf1996} and \citet{gu2003}, for low charge ions. The fits
parameters for bare through Na-like ions are taken from \citet{badnell2006} and for other ions we follow
\citet{dere2009}. Dielectronic recombination rates fits coefficients for H-like through Mg-like ions are taken from
\citet{badnell2006}, \citet[and references therein]{za2003,za2004,za2006}, \citet{colgan2003,colgan2004},
\citet{altun2004,altun2006,altun2007}, \citet{mi2004} and \citet{bautista2007} with updates for S{\textsc{vi}}
\citep{orban2009}, Ne{\textsc{vii}} \citep{orban2008}, Fe{\textsc{viii}}-Fe{\textsc{ix}} \citep{sc2008},
Fe{\textsc{x}}-Fe{\textsc{xi}} \citep{le2009}, Fe{\textsc{xiv}} \citep{sc2006}, Fe{\textsc{xv}}
\citep{lu2007}, Fe{\textsc{xxiii}} \citep{savin2006}. For the remaining ions MAZ1998 data is used.

Charge-exchange recombination (CER) with H{\textsc{i}} rates for He{\textsc{ii}}-He{\textsc{iii}},
N{\textsc{ii}}-N{\textsc{v}}, O{\textsc{iii}}, O{\textsc{v}}, Ne{\textsc{iii}}-Ne{\textsc{v}},
Mg{\textsc{iii}}-Mg{\textsc{v}}, Si{\textsc{v}}, S{\textsc{iii}}-S{\textsc{v}} and Fe{\textsc{iii}}-Fe{\textsc{v}} are
taken from \citet{kf1996}, C{\textsc{ii}}-C{\textsc{vii}} \citep{suno2006}, O{\textsc{ii}} \citep{spirko2003},
O{\textsc{iv}} \citep{wang2003}, Si{\textsc{iii}} \citep{clarke1998}, and Si{\textsc{iv}} \citep{bruhns2008}. Fits to
the rates of CER with He{\textsc{i}} for N{\textsc{iii}}-N{\textsc{v}}, O{\textsc{iii}}, O{\textsc{v}},
Ne{\textsc{iv}}-Ne{\textsc{v}}, C{\textsc{iv}}-C{\textsc{v}}, Mg{\textsc{iv}}-Mg{\textsc{v}},
Si{\textsc{iv}}-Si{\textsc{v}}, S{\textsc{iv}} and Fe{\textsc{iv}}-Fe{\textsc{v}} are taken from
Astrophysics Charge-Transfer Database \citep[and references therein]{wang2002a}, O{\textsc{ii}}
\citep{zhao2005a}, O{\textsc{iv}} \citep{wu2009}, Ne{\textsc{iii}} \citep{zhao2006}, S{\textsc{iii}} \citep{zhao2005b}, 
S{\textsc{v}} \citep{wang2002}, and Fe{\textsc{vi}}-Fe{\textsc{xiv}} \citep{cadez2003}. Charge-exchange ionization (CEI)
with H{\textsc{ii}} rates for Mg{\textsc{ii}}, Si{\textsc{i}}-Si{\textsc{ii}} are taken from AR1985, C{\textsc{i}},
N{\textsc{i}}, Mg{\textsc{i}}, S{\textsc{i}} and Fe{\textsc{i}}-Fe{\textsc{ii}} are taken from \citet{kf1996} and
O{\textsc{i}} from \citet{spirko2003}. CEI with He{\textsc{ii}} comprised data for O{\textsc{i}} \citep{zhao2004}
and Si{\textsc{ii}} \citep{wang2002}, C{\textsc{ii}}, N{\textsc{ii}}, Si{\textsc{iii}} and
S{\textsc{ii}}-S{\textsc{iii}} (AR1985).

Cooling rates include free-free emission with the averaged Gaunt factor by Karzas \& Latter (1961), radiative and
dielectronic recombination \citep{ct1969}, line emission in the range 1 \AA~-610 $\mu$
\citep{penston1970,jd1972,dm1972,kato1976,stern1978,gaetz1983,mewe1985}, and two-photon emission. Spectra calculations
include line and continuum emissions. The latter comprises free-free, free-bound to the ground and excited states, and
two-photon - using 1s-2s transitions in H and He-like ions \citep{tg1966,gr1978,mewe1986}.

\subsection{Equations}

The time evolution of the ions fractions, where ionization and recombinations of ions of nuclear charge $Z$ occur
between neighbouring ionization stages $z-1$, $z$ and $z+1$, is given by
\begin{equation}
\frac{d n_{Z,z}}{d t} ={\cal
I}_{Z,z-1}n_{Z,z-1}n_{e}-({\cal I}_{Z,z}+ {\cal R}_{Z,z})n_{Z,z}n_{e}+{\cal
R}_{Z,z+1}n_{Z,z+1}n_{e},
\end{equation}
where ${\cal R}_{Z,z}$ and ${\cal I}_{Z,z}$ are the rates of recombination and ionization from state $(Z,z)$ to
$(Z,z-1)$ and $(Z,z+1)$, respectively; $n_{Z,z}$ and $n_{e}$ are the ion density of element $Z$ with effective charge
$z$ and electron density, respectively. This constitutes a tridiagonal matrix if the charge exchange reactions are not
included. However, if charge exchange reactions are included (as done in the present paper) new off-diagonals terms are
introduced, becoming 
\begin{equation}
{\cal R}_{Z,z}=\alpha^{r}_{Z,z}+\alpha^{d}_{Z,z}+\frac{1}{n_{e}}\sum \alpha^{ce}
\tilde{n}_{\tilde{Z},\tilde{z}},
\end{equation}
where the uppers indices stand for radiative, dielectronic and charge exchange processes,
respectively, and
\begin{equation}
{\cal I}_{Z,z}=C^{eii}_{Z,z}+C^{ea}_{Z,z}+\frac{1}{n_{e}}\sum C^{ce}\tilde{n}_{\tilde{Z},\tilde{z}}
\end{equation}
with $eii$ and $ea$ standing for electron impact and excitation-autoionization processes, respectively, and
$\tilde{n}_{\tilde{Z},\tilde{z}}$ is the particle density of other ions involved in the charge exchange
reactions. In the EPEC version referred to here, we do neither consider photoionization nor
ionization due to suprathermal electrons \citep[see, e.g.,][]{st1993} - this is the
subject of a forthcoming paper. For the 10 elements considered in the EPEC, the number of ordinary
differential equations including the neutrals amounts to 112. Conservation of species (atoms and ions) 
implies that
\begin{equation}
 n_{Z}=\sum_{z=0}^{Z}n_{Z,z}.
\end{equation}
In addition the system of equations must account for the mass
\begin{equation}
 n_{tot}=\sum_{Z}n_{Z}+n_{e}
\end{equation}
and charge conservations
\begin{equation}
n_{e}=\sum_{Z}\sum_{z=1}^{Z}n_{Z,z}z
\end{equation}
The system of equations is closed by the energy balance equation
\begin{equation}
\label{energyeq}
\frac{d U}{dt}-\frac{P}{n} \frac{d n}{dt}=-n_{e}n_{H}\Lambda
\end{equation}
where $\Lambda$ is the cooling function, i.e., it represents the radiative losses per unit emission
measure resulting from bremsstrahlung, radiative and dielectronic recombination, collisional
ionization, line emission due to excitation, two-photon emission, and charge-exchange reactions. In
these equations $n_{H}$ is the hydrogen density and $U$ is the internal energy density of the
system. The internal energy comprises the contributions from thermal motions of the particles, represented by $U_{th}$,
and the potential energy associated with the ionization stages, that is the energy stored in or delivered from the high
ionization stages of chemical elements \citep{st1993}. Hence, the internal energy density is given by
\begin{equation}
U=U_{th}+\sum_{Z}\sum_{z=1}^{Z}\left(n_{Z,z}\sum_{z^{\prime}=0}^{
z-1}I_{Z,z^{\prime }} \right),
\end{equation}
where $I_{Z,z^{\prime}}$ is the ionization potential of an ion with nuclear charge $Z$ and effective charge
$z^{\prime}$. The thermal part of the internal energy is linked to the pressure of the system through the
equation of state $P=(\gamma -1) U_{th}$.

As initial conditions for static plasma calculations we assume a fully ionized gas at $10^{9}$ K. As the temperature
decreases, recombination and ionization rates are calculated, following the simultaneous implicit calculation of the
ionization fractions (including neutrals) and charge equation -- implying the inversion of a $113 \times 113$ matrix.
Next the radiative losses, emission spectra and internal energy are calculated. 

\section{Collisional ionization Equilibrium vs. Non-Equilibrium ionization}

\begin{figure}[!hb]
\centering
\includegraphics[width=0.6\hsize,angle=0]{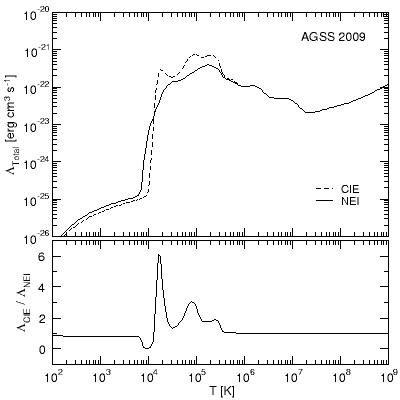}
\caption{\emph{Top panel}: Normalized CIE (dashed line) and NEI (solid line) cooling functions as
function of temperature and calculated with AGSS2009 solar abundances. \emph{Bottom panel}: CIE and
NEI cooling functions ratio.}
\label{cooling}
\end{figure}

\begin{figure}[!ht]
\centering
\includegraphics[width=\hsize,angle=0]{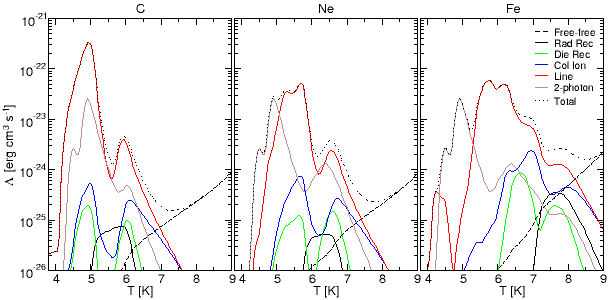}
\includegraphics[width=\hsize,angle=0]{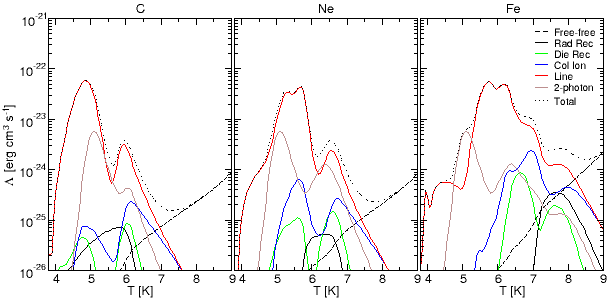}
\caption{Contributions to the CIE (top row) and NEI (bottom row) cooling due to bremsstrahlung
(black dashed line), radiative and dielectronic recombination (black and green solid lines,
respectively), collisional ionization (blue line), line excitation (red line) and two-photon
emission (brown line).}
\label{processes}
\end{figure}

Figure~\ref{cooling} shows the normalized cooling function of a static plasma (that is, with no dynamics included) that
cooled from $10^{9}$ to $10^{2}$ K under CIE and NEI (isochorically) conditions calculated with EPEC and using AGSS2009
abundances.
Variations by factors of 6 at $10^{4.2}$ K are observed between the two efficiencies with
the CIE cooling dominating over the NEI case between $10^{3.6}$ K and $10^{6}$ K (bottom panel of
Figure~\ref{cooling}). These variations are a consequence of the cooling efficiency due to the
different emission processes (bremsstrahlung, radiative and dielectronic recombination, collisional
ionization, line excitation and two-photon emission) being larger under CIE than NEI conditions for
all elements taken into account. Figure~\ref{processes} compares the contributions of these
processes to the cooling per element (e.g, C, Ne and Fe) under CIE (top row) and NEI (bottom row)
conditions.

\begin{figure}[thbp]
\centering
\includegraphics[width=0.5\hsize,angle=0]{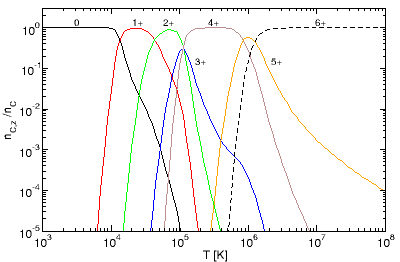}\includegraphics[width=0.5\hsize,angle=0]
{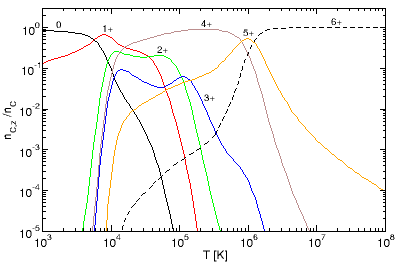}
\includegraphics[width=0.5\hsize,angle=0]{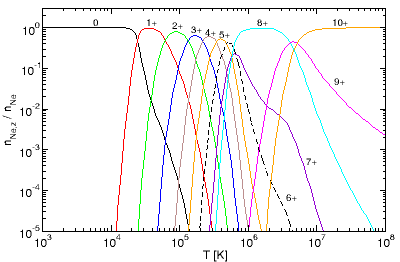}\includegraphics[width=0.5\hsize,angle=0]
{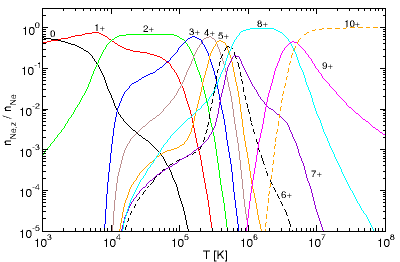}
\caption{Temperature variation of the ionization structure of C (top) and Ne (Bottom) under CIE
(left column) and NEI (right column) conditions for a gas cooling from $10^{9}$ K.}
\label{ionsfracs}
\end{figure}

For T$ <10^{4.1}$ K the cooling efficiency at the same temperature under NEI (isochoric) conditions is larger than in
the CIE case as a result of the delayed recombination of the plasma. As recombination lags behind, single and double
ionized species exist at lower temperatures (right column in Figure~\ref{ionsfracs}), something that does not occur
under CIE, because ionization and recombination is synchronized, and therefore neutrals form at temperatures near
$10^{4}$ K (left column in Figure~\ref{ionsfracs}).

While in CIE  the ionization fractions (left panel Figure~\ref{ionsfracs}) depend only on the temperature and are
sharply peaked, in NEI these same fractions (right panel Figure~\ref{ionsfracs}) depend on the dynamical and thermal
history of the plasma. The higher ionization stages recombine to lower ones and eventually (when T$\sim 10^{3.8}$ K)
only the lowest stages are abundant. However, the qualitative behaviour of all ionic stages is not the same. The highest
ionic stages decrease continuously, that is, they always recombine to the next lowest stages, while the lowest stages
increase continuously, that is the next highest stage recombines to them; the intermediate stages have two peaks
resulting from the recombination of the next highest stage, but dielectronic recombination rapidly depletes it, leading
to the formation of the next lower ionization stage. When dielectronic recombination is no longer effective, 
recombination from the next highest stage increases the ion amount. As soon as the highest stage is
depleted, the next stage recombines to the next lowest stage. 

\begin{figure}[!ht]
\centering
\includegraphics[width=0.5\hsize,angle=0]{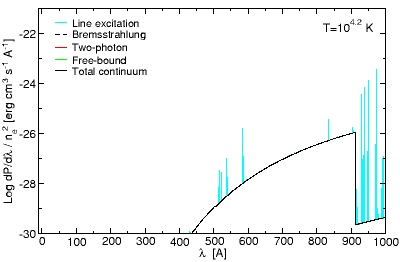}\includegraphics[width=0.5\hsize,angle=0]
{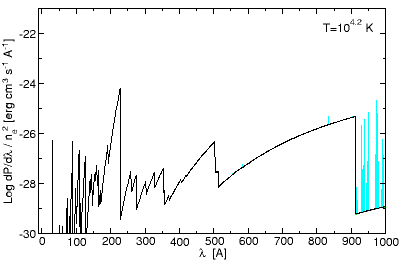}
\includegraphics[width=0.5\hsize,angle=0]{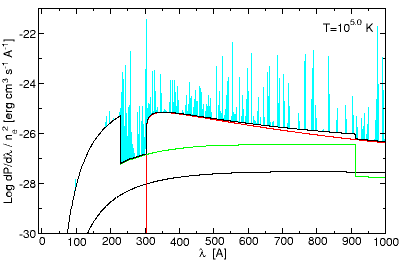}\includegraphics[width=0.5\hsize,angle=0]
{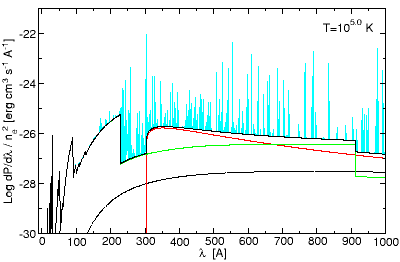}
\caption{Spectra of a plasma that cooled from $10^{9}$ K under CIE (left column) and NEI
(right column) conditions at $10^{4.2}$ K (top panel) and $10^{5}$ K (bottom panel). The total
continuum (black solid line) results from free-free (black dashed line), free-bound (green line) and
two-photon (red line) emissions. Line emission is shown in cyan for the different wavelengths. Note 
the striking differences between CIE and NEI emission spectra.}
\label{spectra}
\end{figure}

These differences between CIE and NEI become quite noticeable in the emission spectra of the plasma at different
temperatures (Figure~\ref{spectra}). The spectra include line emission (cyan lines) and continuum (free-free (dashed
black line), free-bound (green line) and two-photon (red line) emissions). With the decrease in temperature from
$10^{6}$ to $10^{4.2}$ the CIE and NEI spectra become quite different as result of the free-bound emission dominating
the spectra up to $500$ \AA~ at low temperatures. At high temperatures, above $10^{6}$ K the differences between the
ionization structure under CIE and NEI conditions are small, and therefore the spectra
in these two cases are similar. They differ appreciably when the recombination lags behind, typically  around $10^{5}$ K, with the continuum at low wavelengths
(i.e., $\lambda < 100$\AA) becoming free-bound dominated.
Note that in the CIE case there is very little ($\ll 10^{-30}$ erg cm$^{-3}$ s$^{-1}$ $\AA^{-1}$) emission with
increasing range of wavelengths and decrease in temperature, e.g, at $10^{5}$ K: $\lambda<73 \AA$, $10^{4.6}$ K:
$\lambda<179 \AA$ and $10^{4.2}$ K: $\lambda<435 \AA$. Hence, no X-ray emission is expected in the latter two cases
under CIE unlike in NEI, where due to delayed recombination, there is emission to be expected in these ranges of
wavelengths and temperatures (right panel Figure~\ref{spectra}). 

\section{Signature of the Initial Conditions in the Turbulent ISM - NEI Modelling}

In a turbulent supernova-driven ISM the state of the plasma, and therefore its ionization structure, is determined by
the heating and cooling as well as by the flow dynamics. Hence, it is expected that the ionization structure of the
plasma varies from place to place leading to a multitude of cooling functions in the computational box. 
\begin{figure}[!h]
\centering
\includegraphics[width=0.48\hsize,angle=0]{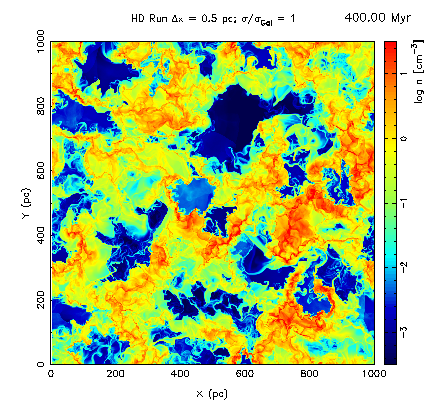}\includegraphics[width=0.5\hsize,angle=0]
{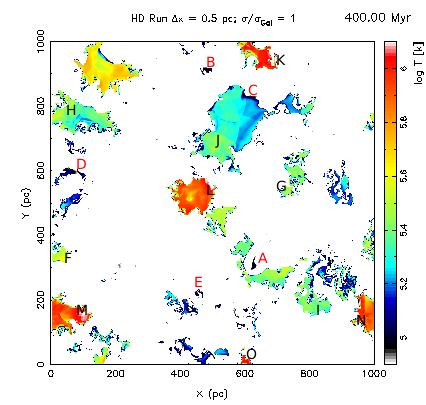}
\caption{\emph{Left panel:} Non-equilibrium ionization density distribution in the Galactic midplane at evolution time
400 Myr. The color scale refers to the logarithm of the number density. Cold (high density) gas is represented by red
while hot (low density) gas is shown in blue. \emph{Right panel:} Midplane gas temperature between $10^{4.9}$ and
$10^{6.1}$ K at 400 Myr of evolution. The labels refer to a selection of sites having temperatures of $10^{5}$ K (labels
A through E), $10^{5.5}$ K (labels F through J) and $10^{6}$ K (labels K through O).}
\label{fig1}
\end{figure}

We carried out three-dimensional hydrodynamical disk-halo interaction simulations (in a patch of the Galaxy located at
the solar radius with an area of 1 kpc$^{2}$ parallel to the Galactic midplane, and extending to $\pm10$ kpc
perpendicular to it, similar to those described in \citet{ab2007} using a resolution of 0.5 pc (the highest so far used
for large scale ISM evolution) corresponding to an effective grid with $2000^{3}$ cells per kpc$^{3}$ boxes) and
including (i) local self-gravity, (ii) heat conduction and (iii) time-dependent evolution of the ionization structure
(of H, He, C, N, O, Ne, Mg, Si, S, and Fe) at each cell of the grid using EPEC described above; (iv) the revised solar
abundances by \citet{agss2009} are used; (v) Periodic boundary conditions are used along the vertical faces, while
outflow boundary conditions are set at the top and bottom of the grid.

The left panel of Figure~\ref{fig1} displays the density distribution in the Galactic midplane at evolution time 400
Myr, while the right panel shows the regions with temperatures between $10^{4.9}$ K and $10^{6.1}$ K. The displayed time
is long enough for (i) the signature of the initial plasma conditions at $t=0$ to be wiped out \citep[it takes some 80
Myr for this to happen;][]{ab2004}, (ii) the disk-halo-disk circulation to be fully established, and (iii) the system to
reach dynamical equilibrium in a statistical sense \citep{av2000,ab2004}. The figure shows in great detail the topology
of the frothy and turbulent ISM permeated with hot gas and having low temperature filamentary structures. 

Bubbles and superbubbles dominate the landscape, but the volume filling factor (that is, the fraction of disk volume) of
the hot gas is merely $\simeq 20\%$ for the Galactic supernova rate. Dark blue regions have temperatures $\geq 10^{7}$ K
resulting from recent supernova activity, whereas darkest red regions are molecular clouds with an excess density of $>
1000$ cm$^{-3}$ and temperature T$<100$ K. Cold gas (red regions) is formed as a result of shock compressed layers and
cooling instabilities in the flow. The details of these simulations due to the high resolution allow us to identify
turbulent motions and their effects at small scales. Superbubbles and bubbles have turbulent flows of material as can be
seen in the cavities, which are crossed by sheet like structures with variable geometries. These turbulent flows are
responsible for the redistribution of energy inside and outside of the cavities.

\begin{figure}[!hb]
\centering
\includegraphics[width=0.32\hsize,angle=0]{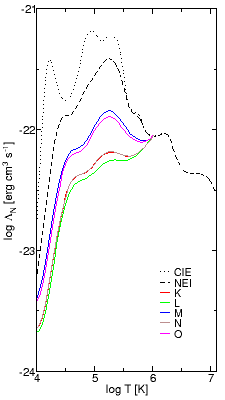}\includegraphics[width=0.68\hsize,angle=0]
{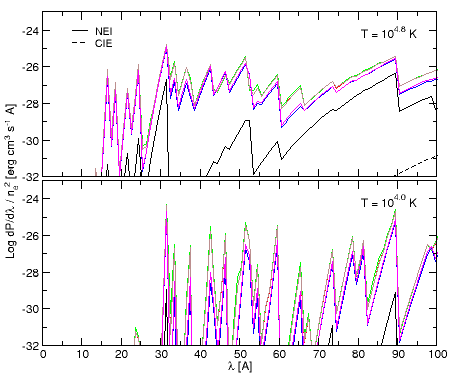}
\caption{\emph{Left panel:} Normalized cooling functions at sites K through O shown in the right panel of
Figure~\ref{fig1} having an initial temperature of $10^{6}$ K. \emph{Right panels:} Free-bound emission (color curves)
from gas having an initial temperature of $10^{6}$ K, located at sites K-O and having the cooling curves displayed in
the left panel. Solid and dashed black lines represent the emission expected from NEI and CIE static (i.e., no dynamics)
plasma at the temperatures shown in each panel.}
\label{spectrarun4}
\end{figure}

Because the plasma keeps a record of its history, and therefore, of its initial condition (temperature, total pressure
and density - including mass fluxes from neighbouring locations), the ionization structure of the plasma varies from
place to place leading to a multitude of cooling functions for the same initial temperature as can be seen in the left
panel of Figure~\ref{spectrarun4}, which displays the cooling paths for gas having an initial temperature of $10^{6}$ K
and located at sites K through O, respectively (shown in the right panel of Figure~\ref{fig1}). The cooling efficiency
fails to match that calculated under NEI (isochoric) conditions (black dashed line) all the way down to $10^{2}$ K (left
panel of Figure~\ref{spectrarun4}), because the latter has less "potential energy" stored in high ionization stages.

An important consequence of this variability seen in the ionization structure evolution and cooling paths is
the occurrence of X-ray emission, through free-bound transitions, at low temperatures, becoming larger than the
corresponding NEI emission in a static plasma. The NEI spectra were calculated for a gas cooling in a time-dependent
fashion from $10^{9}$ K. The right panel of Figure~\ref{spectrarun4} shows that the free-bound emission from the K
through O sites (red, green, blue, brown and magenta lines in the left panel) dominates the NEI (solid black line) and
CIE (dashed black line) static plasma emission with decreasing temperature. This is a clear indication that
recombination from the continuum is not following the cooling of the gas. 

\section{Discussion and Final Remarks}

In this paper we emphasize the importance in ISM simulations to follow the ionization structure of a plasma in a
time-dependent fashion, coupled self-consistently to the dynamics. In fact, the circulation of gas between the disk and
halo is a dynamic process, which involves a time scale, that can be much shorter than any of the microphysical time
scales due to ionization and recombination. The gas escaping into the halo has an initial temperature well in excess of
$10^{6}$ K, where the assumption of collisional ionization equilibrium (CIE) is approximately valid. As the hot plasma
expands away from the disk it will cool adiabatically thereby reducing its temperature and density. The recombination of
highly ionized species lags behind and occurs mainly at considerable heights from the disk. 

New state of the art simulations of the ISM, focusing on the detailed description of the plasma ionization structure in
a supernova driven ISM show that: (i) in a dynamic ISM, the ionization structure and, therefore, the cooling function,
varies with time and from place to place, depending on the initial conditions and its history (a result in accordance to
previous discussions by \citet{kafatos1973}, \citet{shapiro1976} and \citet{sutherland1993} regarding static plasmas,
i.e., with no dynamics included), (ii) the cooling path can be quite different for gas even with the same initial
temperature, but having different densities and pressures, (iii) this path may not follow the one predicted by the pure
plasma emission calculations, that is, without the dynamics included, and (iv) as a consequence, X-ray emission occurs
at temperatures $<10^{5}$ K. This is a consequence of the important emission contribution of \emph{delayed
recombination}, arising from a severe mismatch of recombination and dynamical time scales of the plasma. Hence, as $T$
decreases, the emissivity becomes much larger, by more than an order of magnitude, than that predicted by CIE at the
same temperature.

A fundamental consequence of the previous discussion, is that the cooling at any point in the ISM depends clearly on the
history and dynamics of the plasma, that is, the ionization structure keeps a record of its initial conditions and the
thermodynamic path it has taken. As the ionization structure varies from place to place there exists a multitude of
different cooling functions in the ISM. Furthermore, with delayed recombination having an important role in the cooling
of \emph{overionized} gas, i.e. gas which once had a high temperature during its history, X-ray emission at low
temperatures is expected. Consequently, the mismatch of observed plasmas, e.g. by DXS \citep{sanders2001}, XQC
\citep{mccammon2002}, EUVE \citep{jelinsky1995}, with standard CIE emission models (even for multi-temperature fits) may
eventually result from not correctly describing the NEI structure of the plasmas.

In addition, the amount of singly ionized ions which are important for molecular cloud chemistry, will also be affected
by the timescales for recombination and cooling paths. This implies that a careful study of the ionization structure at
higher temperatures is needed when one is dealing with molecular chemistry in the ISM.

\acknowledgements 

M.A. would like to thank the Portuguese-American Foundation for Development (FLAD) for the financial support, under
project F-V-162/2010, to attend the meeting. This research has been funded by the Foundation for Science and Technology
(Portugal) under project PTDC/CTE-AST/70877/2006 (``The ionization of Diffuse Extraplanar Gas Layers in Spiral
Galaxies'').

\bibliography{aspauthor}

\end{document}